\documentstyle[12pt]{article}

\font\mybb=msbm10 at 11pt
\def\bb#1{\hbox{\mybb#1}}
\def\e{{\rm e}}
\def\slash{\!\!\!/}

\newcommand{\newsection}[1]
{\vspace{5mm}
\pagebreak[3]
\addtocounter{section}{1}
\setcounter{equation}{0}
\setcounter{subsection}{0}
\setcounter{footnote}{0}
\begin{flushleft}
{\large\bf \thesection. #1}
\end{flushleft}
\nopagebreak
\medskip
\nopagebreak}

\newlength{\extraspace}
\newlength{\extraspaces}
\begin{document}

\addtolength{\baselineskip}{.8mm}

\thispagestyle{empty}

\begin{flushright}
{\sc OUTP}-97-08P\\
  hep-th/9702088\\
\hfill{  }\\ \today
\end{flushright}
\vspace{.3cm}

\begin{center}
{\large\sc{Dynamical Description of Spectral Flow in $N=2$
Superconformal Field Theories}}\\[15mm]

{\sc  Leith Cooper, Ian I.\ Kogan and Richard J.\ Szabo
} \\[2mm]
{\it Department of Physics -- Theoretical Physics, University of Oxford\\ 1
Keble Road, Oxford OX1 3NP, U.K.} \\[15mm]

\vskip 1.5 in

{\sc Abstract}

\begin{center}
\begin{minipage}{14cm}

We show how the spectral flow between the Neveu-Schwarz and
Ramond sectors of $N=2$ superconformal field theories can
be described in three dimensions in terms of the propagation
of charged particles coupled to a a Chern-Simons gauge theory.
Quantum mechanical mixing between the degenerate Chern-Simons
vacua interpolates between the different boundary conditions of
the two sectors and so provides a
dynamical picture for the {\sc gso}-projection.

\end{minipage}
\end{center}

\end{center}

\noindent

\vfill
\newpage
\pagestyle{plain}
\setcounter{page}{1}
\stepcounter{subsection}
\newsection{Introduction}
\renewcommand{\footnotesize}{\small}

Understanding the structure of the moduli space of two-dimensional conformal
field theories is an important problem in critical string theory. This moduli
space forms the space of string vacua and describes the appropriate stringy
modification of classical general relativity. Particular attention has been
devoted to models with $N=2$ worldsheet supersymmetry, where the solutions to
the string equations of motion hold to all orders of string perturbation theory
and the resulting physics itself possesses $N=1$ spacetime supersymmetry. In
contrast to $N=1$ superconformal field theories, whose Virasoro algebra is
quite similar to that of the non-supersymmetric ($N=0$) models, the moduli
space and algebraic structures of $N=2$ superconformal field theories are
structurally different (see \cite{greene,aspinwall} for concise reviews). For
example, in the $N=2$ models it is possible to smoothly interpolate between
different parameter space descriptions of the same conformal field theory
leading to physically smooth space-time topology changing processes in string
theory. In this Paper we shall study the algebraic property that the
Neveu-Schwarz and Ramond sectors of the $N=2$ super-Virasoro algebra are
connected by the spectral flow \cite{specflow}. We will describe how this
phenomenon appears as a dynamical property of the coupling of charged matter to
a three-dimensional topological field theory.

The connection between two-dimensional conformal field theories and
three-dimensional topological field theories traces back to the seminal paper
by Witten \cite{witten}, where it was shown that the physical state space of
Chern-Simons gauge theory, with action
\begin{equation}
kS_{\tiny CS}^{[G]}[A]=\frac{k}{4\pi}\int_{\cal
M}\mbox{Tr}\left(A\wedge dA+\mbox{$\frac{2}{3}$}A\wedge A\wedge A\right)
\label{csaction}\end{equation}
defined on the three-dimensional spacetime manifold ${\cal M}=\Sigma\times\bb
R^1$, coincides with the finite-dimensional space of conformal blocks of the
Wess-Zumino-Novikov-Witten ({\sc wznw}) model at level $k\in{\bb Z}$ defined on
a compact Riemann surface $\Sigma$. Here $A$ is a gauge connection of a compact
simple Lie group $G$, and the quantum field theory (\ref{csaction}) provides a
three-dimensional description of the current algebra $G_k$ based on $G$ at
level $k$. This relationship was further developed by Moore and Seiberg
\cite{mooreET} who showed that a Chern-Simons theory defined on a
three-dimensional manifold ${\cal M}$ with two-dimensional boundary
${\partial\cal M}$ induces the chiral gauged {\sc wznw} model on $\partial{\cal
M}$
\begin{eqnarray}
kS_{\tiny WZNW}^{+~[G]}[g,\bar A_z]&=&\frac k{4\pi}\int_{\partial{\cal
M}}d^2z~\mbox{Tr}\left(g^{-1}\partial_zg~g^{-1}\partial_{\bar z}g-2\bar
A_zg^{-1}\partial_{\bar z}g\right)\nonumber\\& &+\frac k{12\pi}\int_{\cal
M}\mbox{Tr}\left(g^{-1}dg\wedge g^{-1}dg\wedge g^{-1}dg\right)
\label{wznwaction}\end{eqnarray}
where $\bar A$ is the bulk part of the gauge field $A$ in $\cal M$ and
$g(z,\bar z)\in G$ is the pure gauge part of $A$ on $\partial\cal M$, i.e.
\begin{equation}
A=g^{-1}\bar Ag+g^{-1}dg
\end{equation}

The key feature of the Moore-Seiberg correspondence is that different algebraic
constructions of conformal field theories on $\partial{\cal M}$ can be
described geometrically by considering combinations of several
Chern-Simons theories in the bulk $\cal
M$. This is done by exploiting the Goddard-Kent-Olive ({\sc gko}) coset
constructions \cite{gko} for representations of the Virasoro algebra. A coset
conformal field theory based on $G_k/H_\ell$, where $H$ is a subgroup of $G$
such that the Sugawara-Virasoro algebra over $G$ decomposes orthogonally into
two mutually commuting Sugawara-Virasoro algebras over $G/H$ and $H$, can be
represented by an action which is the difference $kS_{\tiny WZNW}^{[G]}-\ell
S_{\tiny WZNW}^{[H]}$ of {\sc wznw} actions. Using the above correspondences
this can be described by Chern-Simons gauge theory. For example, the coset
\begin{equation}
M_k=SU(2)_k\times SU(2)_1/SU(2)_{k+1}
\label{N=0minimal}\end{equation}
represents the ordinary $N=0$ minimal models of the Virasoro algebra in terms
of the affine Kac-Moody algebra $SU(2)_\ell$ based on the group $SU(2)$ at
level $\ell$. It can be described using three independent $SU(2)$ Chern-Simons
gauge fields with the action
\begin{equation}
{\cal I}^{\{N=0\}}[A,B,C]=kS_{\tiny CS}^{[SU(2)]}[A]+S_{\tiny
CS}^{[SU(2)]}[B]-(k+1)S_{\tiny CS}^{[SU(2)]}[C]\,.
\end{equation}
Analogously, the $N=1$ superconformal minimal
models are described by the coset
\begin{equation}
{\cal S}M_k={\cal S}SU(2)_k\times{\cal S}SU(2)_2/{\cal S}SU(2)_{k+2}
\label{N=1minimal}\end{equation}
of the $N=1$ supersymmetric Kac-Moody algebra ${\cal S}SU(2)_\ell$ based on
$SU(2)$ at level $\ell$, and they can be obtained from the $N=1$
supersymmetric $SU(2)$ Chern-Simons theory \cite{sakai} with action
\begin{equation}
{\cal I}^{\{N=1\}}[A,B,C]=kS_{\tiny SUSY~CS}^{[SU(2)]}[A] + 2S_{\tiny
SUSY~CS}^{[SU(2)]}[B]-(k+2)S_{\tiny SUSY~CS}^{[SU(2)]}[C]\, .
\end{equation}

The three-dimensional approach is a good tool in the classification of rational
conformal field theories \cite{mooreET} because it provides a
geometrical realization of the two-dimensional models using
conventional techniques of quantum gauge theory. It has been shown
recently \cite{kogan1} that the coupling of the Chern-Simons gauge 
theory to dynamical
charged matter fields has dramatic consequences. First of all, the field theory
becomes dynamical, except in the low-energy limit where it can still be
regarded as some sort of topological field theory. At the quantum level, the
dynamical charged matter fields will induce a Yang-Mills kinetic term for the
gauge field and will also add dynamical degrees of freedom in the
gauge sector of
the theory. This, as we shall see in this Paper, leads to very important
dynamical effects which are responsible for the occurence of some fundamental
properties of the induced conformal field theory. Another very important
feature of a matter coupling is that charged scalar matter can describe a
deformation of the respective conformal field theory. This means that some
specific $N=2$ superconformal models that we shall study can be considered as
special points in the moduli space of $N=2$ superconformal field theories and
other $N=2$ models are related to them by marginal deformations which can be
obtained in a three-dimensional description by varying the parameters of the
charged matter (for example the chemical potentials). Thus the addition of
charged matter to the theory could be related to some of the exotic properties
that the $N=2$ models possess, such as mirror symmetry and spacetime topology
change \cite{greene,aspinwall}. The detailed description of these ideas for the
$N=0$ and $N=1$ models has been presented in \cite{kogan1} where the
transitions between deformed minimal models using the above three-dimensional
descriptions were studied.

In the following we shall study three-dimensional constructions for $N=2$
superconformal field theories. We will use the fact \cite{szaboET} that in the
$N=0$ and $N=1$ models the fundamental observables of the conformal field
theory, i.e. the anomalous scaling dimensions of primary operators, correspond
to the transmuted
spins that appear as Aharonov-Bohm phases from adiabatical rotation of charged
particles coupled to Chern-Simons gauge fields. We shall show how the basic
observables of the $N=2$ models can be similarly described and demonstrate that
this three-dimensional representation yields geometric and dynamical
realizations of the spectral flow between isomorphic $N=2$ superconformal
algebras. When the spectral flow interpolates between the Neveu-Schwarz and
Ramond sectors, we will see that an appropriate combination of basis states in
the Hilbert space of the Chern-Simons gauge theory on ${\cal
M}=\Sigma\times{\bb R}^1$ (corresponding to a choice of spin structure for the
fermion fields on the compact Riemann surface $\Sigma$) coincides with the
necessary truncation of the world-sheet spectrum required for the
superconformal field theory to further possess $N=1$ spacetime supersymmetry,
i.e. the Gliozzi-Scherk-Olive ({\sc gso}) projection \cite{gso}. The physical
interpretation thus obtained is in terms of the quantum mechanical Landau
problem. These results imply some intriguing target space properties of the
so-called topological membrane approach to string theory \cite{tm}, which
predominantly describes
world-sheet modifications of string theory by filling in the string world-sheet
and viewing it as the boundary of a 3-manifold. In this way, the
three-dimensional description that we present here suggests dynamical and
geometric origins for the appearence of space-time supersymmetry in string
theory.

The structure of the remainder of this Paper is as follows. In Section 2 we
discuss some basic properties of the $N=2$ super-Virasoro algebra and introduce
the coset models that will be studied. In Section 3 we show how these coset
models can be described using three-dimensional topological field theory
coupled to sources and show that this description leads immediately to a
physical interpretation of the spectral flow. In Section 4 we construct the
Hilbert space of the relevant Chern-Simons gauge theory in the canonical
formalism and show that the vacuum sector admits a choice of basis states
appropriate to the various spin structures of the superpartner spinor fields on
$\Sigma$, and hence to the {\sc gso} projection and also modular invariance of
superstring theory. Finally, Section 5 contains some concluding remarks as well
as generalizations of our analysis to other $N=2$ superconformal field
theories.

\newsection{Coset Models for $N=2$ Superconformal Field Theories}

We begin by briefly discussing some aspects of the $N=2$ superconformal algebra
and its coset representations. The $N=2$ superconformal algebra (in the
holomorphic sector of the world-sheet theory) is generated by the usual
Virasoro stress-energy tensor $T(z)$, an extra $U(1)$ current $J(z)$ of
conformal dimension 1 and two supercurrents $G^\pm(z)$ with $U(1)$ charges $\pm
1$ (for the precise relations satisfied by these generators, see
\cite{greene}). The fermionic currents are also labelled by an additional
parameter
$\eta\in(-\frac12,\frac12]$ which controls their boundary conditions as
\begin{equation}
G^\pm(\e^{2\pi i}z)=\e^{\mp 2\pi i(\eta+1/2)}G^\pm(z)
\label{bc}
\end{equation}
In particular, when $\eta=0$, $G^{\pm}(z)$ are anti-periodic giving us the
Neveu-Schwarz sector of the theory, and when $\eta=1/2$, $G^{\pm}(z)$ are
periodic yielding the Ramond sector. At first glance, different choices of
$\eta$ appear to determine different algebras with (slightly) different
commutation relations, but it turns out that the spectral parameter $\eta$ in
fact labels a family of {\it isomorphic} $N=2$ superconformal algebras. We
shall return to this property in the next Section.

We seek a realization of the $N=2$ theories in terms of Chern-Simons gauge
theory. This can be done using the Kazama-Suzuki coset models \cite{ks} which
encompass most of the known $N=2$ superconformal field theories and provide a
coset realization of the $N=2$ super-Virasoro algebra. The Kazama-Suzuki coset
construction of $N=2$ superconformal field theories first applies
the {\sc gko} coset construction with $N=1$ supersymmetric Kac-Moody algebras
to obtain a large class of $N=1$ superconformal models, and then it examines
under which conditions these $N=1$ coset models so constructed also possess an
extended $N=2$ supersymmetry. It turns out that these conditions are met if $G$
is a compact simple Lie group and $H$ is a subgroup of $G$ such that the group
manifold $G/H$ is a Hermitian symmetric space. In that case, the coset
\begin{equation}
M_k^{\tiny(KS)}\equiv{\cal S}G_{k+C_2(G)}/{\cal S}H_{k+C_2(G)}
\cong G_k\times SO(2n)_2/H_{\ell(k)}
\label{ks}
\end{equation}
contains the $N=2$ super-Virasoro algebra, where
\begin{equation}
\ell(k)=k+C_2(G)-C_2(H)
\end{equation}
and $C_2(G)$ (respectively $C_2(H)$) is the dual Coxeter number of the Lie
group $G$ ($H$). The $SO(2n)_2$ part of the coset (\ref{ks}) represents the
internal symmetry group of $2n$ free Majorana (or $n$ free Dirac) fermion
fields, where
$n=\mbox{rank}(G)=\mbox{rank}(H)$. The explicit form of the $N=2$ generators in
terms of these free fermion fields and the Kac-Moody currents of $G_k$ can be
found in \cite{ks}. For instance, the extra $U(1)$ current $J(z)$ can be
represented via the embedding $SO(2)\subset SO(2n)$ in (\ref{ks}).

The isomorphism in (\ref{ks}) between the $N=1$ and $N=0$ cosets can be proven
algebraically \cite{ks}. A more explicit argument can be given by exploiting
the fact that the cosets in (\ref{ks}) determine orthogonal decompositions of
the algebras and thus representing the coset of the $N=1$ supersymmetric
Kac-Moody algebras in (\ref{ks}) in terms of {\sc wznw} actions. The crucial
feature of these locally invariant actions is that the
superpartner kinetic terms are gauged with respect to the underlying Lie group
of the current algebra. For instance, the relevant kinetic terms in the $N=1$
supersymmetric {\sc wznw} action for the current algebra ${\cal S}G_{k+C_2(G)}$
in (\ref{ks}) can be written as
\begin{eqnarray}
(k+C_2(G))S_{\tiny WZNW}^{[G]}[g]+(k+C_2(G))\int_\Sigma
d^2z~\bar\psi\left(i\partial\slash+g^{-1}\partial\slash
g\right)\psi\nonumber\\~~~~~~~~~~=(k+C_2(G))S_{\tiny WZNW}^{[G]}[g]+(k+C_2(G))
\int_\Sigma d^2z~\bar\psi'~i\partial\slash~\psi'
\label{N=1wznw}\end{eqnarray}
where
\begin{equation}
\psi'(z,\bar z)=g(z,\bar z)\psi(z,\bar z)
\label{psi'g}\end{equation}
Here $\psi$ are Majorana fermion fields with two real spinor components, each
of which is a $(\dim G)$-component field in the adjoint representation of $G$.
The local field transformation (\ref{psi'g}) shows that the fermionic part of
the $N=1$ action (\ref{N=1wznw}) completely decouples and can be written as a
free
action for $\dim G$ Majorana fermion fields. An identical relation holds for
the
current algebra ${\cal S}H_{k+C_2(G)}$ in (\ref{ks}), so that when the coset
${\cal S}G_{k+C_2(G)}/{\cal S}H_{k+C_2(G)}$ is written as the difference of
$N=1$ {\sc wznw} actions (\ref{N=1wznw}) for $G$ and $H$, we are effectively
left with $\dim G-\dim H=\dim(G/H)=2n$ free Majorana fermion fields. Upon
bosonization \cite{witten2}, the action for these $2n$ Fermi fields is
equivalent to the ordinary $N=0$ $SO(2n)_2$ {\sc wznw} model. Thus the
$SO(2n)_2$ current algebra measures the residual reduction of the fermionic
currents between the $N=1$ and $N=0$ cosets based on $G/H$. Moreover, the
Jacobians for the field transformations (\ref{psi'g}) are anomalous and lead to
the levels of the $N=0$ current algebras indicated in (\ref{ks}) by effectively
shifting $k\to k-C_2(G)$ and $k\to k-C_2(H)$ in the coefficients of the bosonic
parts $S_{\tiny WZNW}^{[G]}$ and $S_{\tiny WZNW}^{[H]}$, respectively, of the
actions in (\ref{ks}) \cite{susyiso}.

The $N=2$ superconformal primary fields are labelled (in part) by their usual
conformal dimensions $\Delta$, and also their extra $U(1)$ charges $\cal Q$.
For the Kazama-Suzuki cosets (\ref{ks}), they are given by \cite{ks}
\begin{eqnarray}
\Delta_R &=& \frac{T_R(G)}{k+C_2(G)}-\frac{\lambda^2+2\lambda\cdot\rho_H}
{k+C_2(G)}\, ,\label{dimension}\\
{\cal Q}_R &=& -\frac{4\lambda\cdot(\rho_G-\rho_H)}{k+C_2(G)}\, ,\label{charge}
\end{eqnarray}
in the Neveu-Schwarz sector, where
\begin{equation}
T_R(G)~\delta^{ab}={\rm Tr}~R^aR^b
\end{equation}
is the quadratic Casimir of a unitary irreducible representation $R$ of $G$,
$\lambda$ is a corresponding highest weight vector of $H$, and $\rho_G$ and
$\rho_H$ are the Weyl vectors of $G$ and $H$, respectively. Unitarity of the
highest-weight representation $R$ of the Kac-Moody group $G_k$ imposes the
constraint
\begin{equation}
2\alpha\cdot\Lambda_R/\alpha^2\leq k
\label{wtconstr}\end{equation}
where $\Lambda_R$ is the corresponding highest weight vector and $\alpha$ the
highest root vector of $G$ (along with a similar constraint for the subgroup
$H$). Generally, the Virasoro central charge of the
current algebra $G_\ell$ is $c(G_\ell)=\ell\dim G/\big(\ell+ C_2(G)\big)$, and
$c=1/2$ for each Majorana fermion, so that the central charge of (\ref{ks}) is
\begin{equation}
c(M^{\tiny(KS)}_k)=n+\frac{k\dim G}{k+C_2(G)}-\frac{\ell(k)\dim H}{k+C_2(G)}\,
{}.
\label{central}
\end{equation}

The general Kazama-Suzuki coset models (\ref{ks}) can be described in
three-dimensional terms by combining several Chern-Simons theories in exactly
the same manner as described in Section 1 for the $N=0$ and $N=1$ cases.
However, in this Paper we shall only consider the $N=2$ super-Virasoro algebra
itself which corresponds to $G=SU(2)$ and $H=U(1)$ in (\ref{ks}). This choice
leads to the class of $N=2$ superconformal minimal models
\begin{equation}
{\cal SS}M_k={\cal S}SU(2)_{k+2}/{\cal S}U(1)_{k+2}\cong SU(2)_k\times
SO(2)_2/U(1)_{k+2}
\label{minimal}
\end{equation}
These minimal models are important in their own right because they describe a
particular compact region of the full moduli space of \mbox{$N=2$}
superconformal field theories and one can flow to other $N=2$ models (such as
Calabi-Yau non-linear sigma-models or Landau-Ginsburg theories) by smoothly
varying the moduli space parameters \cite{greene,aspinwall}. Note that the
$N=2$ minimal models (\ref{minimal}) are algebraically much simpler than the
$N=0$ and
$N=1$ minimal models (\ref{N=0minimal}) and (\ref{N=1minimal}). This is one of
the distinguishing features of $N=2$ superconformal field theories and can even
be taken as a motivation for the introduction of higher-supersymmetry (as
higher-$N$ leads to simpler coset constructions). Besides their apparent
simplicity, and the fact that their quantum field algebras have a finite number
of primary conformal fields, these models also have a nice three-dimensional
dynamical interpretation which we shall describe in the next Section.

\newsection{Three-dimensional Description and the Spectral Flow}

{}From (\ref{minimal}) we are thus led to consider the Chern-Simons theory
\begin{equation}
{\cal I}^{\{N=2\}}[A,B,C]=kS_{\tiny CS}^{[SU(2)]}[A]+2S_{\tiny CS}^{[SO(2)]}[B]
-(k+2)S_{\tiny CS}^{[U(1)]}[C]
\label{abc}
\end{equation}
The isomorphism between the $N=1$ and $N=0$ gauge-invariant Chern-Simons
theories in (\ref{minimal}) can be established in much the same way as we did
in the previous Section using {\sc wznw} models, since in the three-dimensional
case the kinetic terms for the $N=1$ fermionic superpartner fields will involve
gauge-covariant derivatives. Now the fact that the $N=0$ $SO(2)$ Chern-Simons
gauge field $B$ in (\ref{abc}) represents the fermionic part of the $N=2$
supersymmetric theory is just the fermion-boson transmutation property of
abelian Chern-Simons gauge theory at level $\ell=2$ \cite{polyakov}, i.e.
that charged particles with Fermi statistics acquire Bose statistics from their
interaction with a level-2 Chern-Simons gauge field (and vice-versa).

We want to describe the basic observables (\ref{dimension}) and (\ref{charge})
of the $N=2$ minimal models (\ref{minimal}) using the action (\ref{abc}). The
first step is to introduce charged matter in such a way that the induced spin
of this charged matter coincides with the conformal dimensions of the minimal
model. Generally, the anomalous spin acquired by charged matter in a
representation $R$ of the gauge group $G$ from its interaction with a
Chern-Simons gauge field at level $\ell$ is given by the Knizhnik-Zamolodchikov
formula \cite{KZ}
\begin{equation}
\Delta_R(G_\ell)=\frac{T_R(G)}{\ell+C_2(G)}
\label{KZspin}\end{equation}
for the anomalous scaling dimensions of primary operators in the corresponding
{\sc wznw} model. The induced spins (\ref{KZspin}) can be interpreted
\cite{anyon} as the Aharonov-Bohm phases that arise from adiabatical rotation
of particles of charge $\sqrt{T_R(G)}$, which also carry an induced magnetic
flux from their interaction with the Chern-Simons gauge field (see
(\ref{flux})), about one another, i.e.
\begin{equation}
\Psi(\e^{2\pi i}(x_1-x_2))=\e^{4\pi i\Delta_R(G_\ell)}~\Psi(x_1-x_2)
\end{equation}
where $\Psi$ is the 2-particle wavefunction. For a perturbative description of
these weights as the Aharonov-Bohm scattering amplitudes between dynamical
charged particles, see \cite{szaboET}.

The conformal dimensions (\ref{KZspin}) can be derived non-perturbatively as
the phase of the determinant of the propagator in the quadratic approximation
to the Chern-Simons action (\ref{csaction}) \cite{witten}. Their equivalence to
the scaling weights in a conformal field theory can be understood geometrically
and dynamically as follows. The $n$-point correlation functions of the
conformal field theory can be decomposed
\begin{equation}
\left\langle\prod_{i=1}^nV^{(R_i)}(z_i,\bar
z_i)\right\rangle=\left\langle\prod_{i=1}^nV_{\tiny
L}^{(R_i)}(z_i)\right\rangle\left\langle\prod_{i=1}^nV_{\tiny R}^{(R_i)}(\bar
z_i)\right\rangle
\label{CFTcorr}\end{equation}
in terms of products of left and right conformal blocks, where $V_{\tiny
L}^{(R_i)}(z_i)$ and $V_{\tiny R}^{(R_i)}(\bar z_i)$ are the holomorphic and
anti-holomorphic chiral vertex operators corresponding to the left-right
symmetric vertex operator $V^{(R_i)}(z_i,\bar z_i)$ in a representation $R_i$
of the group $G$. In the corresponding three-dimensional description we
consider the 3-manifold ${\cal M}=\Sigma\times[0,1]$ whose two boundaries
$\Sigma_{\tiny L}$ and $\Sigma_{\tiny R}$ are connected by a finite interval. A
Chern-Simons gauge theory in $\cal M$ induces both left- and right-moving
sectors of the two-dimensional conformal field theory, and an insertion of a
vertex operator on the worldsheet $\Sigma$ is equivalent to insertions of the
chiral vertex operators $V_{\tiny L}^{(R)}(z)$ and $V_{\tiny R}^{(R)}(\bar z)$
on the left- and right-moving worldsheets $\Sigma_{\tiny L}$ and $\Sigma_{\tiny
R}$, respectively. The insertions corresponding to the correlation functions
(\ref{CFTcorr}) are induced by path-ordered products of the open Wilson line
operators \cite{witten,mooreET,szaboET,tm}
\begin{equation}
W_{{\cal C}_{z_1,\bar z_1};\dots;{\cal C}_{z_n,\bar
z_n}}^{(R_1,\dots,R_n)}[A^{(1)},\dots,A^{(n)}]=\prod_{i=1}^n~\mbox{Tr}~P\exp
\left(i\int_{{\cal C}_{z_i,\bar z_i}}A^{(i)a}R_i^a\right)
\label{wilsonvert}\end{equation}
along the oriented paths ${\cal C}_{z_i,\bar z_i}\subset{\cal M}$ with
endpoints $z_i\in\Sigma_{\tiny L}$ and $\bar z_i\in\Sigma_{\tiny R}$.
Correlators of insertions of the Wilson lines (\ref{wilsonvert}) in $\cal M$
induce phase factors from adiabatical rotation of charged particles coupled to
the Chern-Simons gauge fields $A^{(i)}$ in the representations $R_i$. The
quantum particles propagate along ${\cal C}_{z_i,\bar z_i}$ from left- to
right-moving worldsheets, so that the corresponding linking of the Wilson lines
from the adiabatical rotations in $\cal M$ are equivalent to braidings of the
associated vertex operators on $\Sigma$ whose induced phases are given by
(\ref{KZspin}). A gas of open Wilson lines (\ref{wilsonvert}) describes charged
matter in $\cal M$ corresponding to a deformation of the two-dimensional
conformal field theory \cite{kogan1}.

To this end, we minimally couple the $SU(2)$ Chern-Simons gauge field $A$ in
(\ref{abc}) to a matter current $J_a^{(j)\mu}R^{(j)a}$ in a spin-$j$
representation $R^{(j)}$ of $SU(2)$, i.e. we add the term $\int_{\cal M}2~{\rm
Tr}~A_\mu^aR^{(j)a}J_b^{(j)\mu}R^{(j)b}$ to the action (\ref{abc}). Then the
induced spin of this charged matter is $j(j+1)/(k+2)$. We also minimally couple
the $U(1)$ Chern-Simons gauge field $C$ at level $-(k+2)$ to a current
$J^{(m)\mu}$ carrying an abelian charge $q=m$, corresponding to the magnetic
quantum numbers of this same spin-$j$ representation of $SU(2)$, by adding the
term $\int_{\cal M}C_\mu J^{(m)\mu}$ to (\ref{abc}). Then the total induced
spin of the matter-coupled action
\begin{equation}
{\cal I}_M^{(j,m)}={\cal I}^{\{N=2\}}[A,B,C]+\int_{\cal
M}\left(2j(j+1)A_\mu^aJ_a^{(j)\mu}+C_\mu J^{(m)\mu}\right)
\label{Imatter}\end{equation}
is
\begin{equation}
\Delta_{j,m}=\frac{j(j+1)}{k+2}-\frac{m^2}{k+2}\, .
\label{spin}
\end{equation}
Next, we consider the Gauss law for the matter-coupled abelian Chern-Simons
gauge field $C$ at level $-(k+2)$,
\begin{equation}
-\frac{(k+2)}{4\pi}\,\epsilon^{0ij}\partial_iC_j=J^{(m)0}\, ,
\label{gauss}
\end{equation}
which follows from varying the action (\ref{Imatter}) with respect to the
temporal component $C_0$ of $C$. It implies that a particle of charge $m$ also
carries magnetic flux
\begin{equation}
\Phi_m\equiv\frac1{2\pi}\int_Dd^2x~
\epsilon^{0ij}\partial_iC_j=-\frac{2m}{k+2}\, .
\label{flux}
\end{equation}
where $D\subset\cal M$ is a disc in a neighbourhood of the charged matter.

The induced spin $\Delta_{j,m}$ and abelian magnetic flux $\Phi_m$ match
precisely the conformal dimensions (\ref{dimension}) and $U(1)$ charges
(\ref{charge}), respectively, of the $N=2$ minimal model ${\cal SS}M_k$ for the
spin-$j$ representation. Note that the highest-weight constraint
(\ref{wtconstr}) in this case is $2|m|\leq2j\leq k$, which for a given level
$k\in{\bb Z}$ of the $SU(2)$ Chern-Simons gauge field $A$ gives the
(finitely-many) allowed values
\begin{equation}
\mbox{$j=0,\frac12,1,\frac32,\dots,\frac k2$}~~~~~;~~~~~m=-j,-j+1,\dots,j-1,j
\end{equation}
for the spin and magnetic quantum numbers of the external charged matter. Thus,
by introducing matter which is charged with respect to the Chern-Simons gauge
fields $A$ and $C$, i.e. considering the matter-coupled Chern-Simons gauge
theory (\ref{Imatter}), we can recover all of the quantum numbers which label
the primary fields of ${\cal SS}M_k$. Moreover, the appearence of the extra
$U(1)$ current $J(z)$ in the extension of the Virasoro algebra to include $N=2$
supersymmetry is, from the three-dimensional viewpoint, manifested as a $U(1)$
Gauss' law enabling the possibility to now measure an {\it abelian} flux
of charged matter coupled to the gauge theory (which is responsible for the
abelian Aharonov-Bohm phases that appear).

At first sight, it appears as though we have obtained a full three-dimensional
description of $N=2$ minimal models without any reference to the $SO(2)$
Chern-Simons gauge field $B$ in (\ref{abc}). As we shall now show, the
inclusion of an $SO(2)$ gauge field is crucial in this three-dimensional
construction and it gives rise to a dynamical picture for the spectral flow
\cite{specflow} between the Neveu-Schwarz and Ramond sectors of $N=2$ minimal
models. To see this, we consider another abelian charge $q=-\eta$. It couples
minimally to the abelian gauge fields $B$ and $C$, so that the action
(\ref{Imatter}) is now modified to
\begin{equation}
{\cal I}^{(\eta)}_M={\cal I}_M^{(j,m)}+\int_{\cal
M}\left(B_\mu+C_\mu\right)J^{(-\eta)\mu}
\label{Ieta}\end{equation}
Particles carrying this charge will acquire an induced spin
$\Delta(SO(2)_2)=\eta^2/2$ from their interaction with $B$,
and $\Delta(U(1)_{-k-2})=-\eta^2/(k+2)$ from their
interaction with $C$. Since the fundamental abelian charge $q=m$ of the minimal
model and the extra charge $q=-\eta$ can interact via exchange of the gauge
field $C$, there is an extra contribution to the Aharonov-Bohm phase from the
adiabatic transport of the charge $q=-\eta$ around the magnetic flux $\Phi_m$
of $q=m$. This yields an additional contribution $\Delta=-\eta\Phi_m$ to the
induced spin, and hence the total induced spin of the matter-coupled action
(\ref{Ieta}) is
\begin{equation}
\Delta^{(\eta)}=\Delta_{j,m}+\frac{\eta^2}{2}-\frac{\eta^2}{k+2}
-\eta\Phi_m=\Delta_{j,m}-\eta\Phi_m+\frac{\eta^2}{6}c_k\,,
\label{nweight}
\end{equation}
where $c_k=3k/(k+2)$ is the Virasoro central charge (\ref{central}) of ${\cal
SS}M_k$. The Gauss law (\ref{gauss}) implies that particles
carrying this extra abelian charge $q=-\eta$ will also carry magnetic
flux $\Phi(SO(2)_2)=-\eta$ due to their interaction with $B$ and
$\Phi(U(1)_{-k-2})=2\eta/(k+2)$ due to their interaction with $C$.
The total flux carried by these particles is thus
\begin{equation}
\Phi^{(\eta)}=\Phi_m-\eta+\frac{2\eta}{k+2}=\Phi_m - \frac{\eta}{3}c_k\,.
\label{ncharge}
\end{equation}

We immediately recognize $\Delta^{(\eta)}$ and $\Phi^{(\eta)}$ as the conformal
weight and $U(1)$ charge of the $\eta$-twisted sector of the
$N=2$ superconformal models \cite{greene,specflow}. The three-dimensional
abelian charge $\eta$ is identified with the spectral parameter appearing in
(\ref{bc}) which interpolates between the different boundary conditions on the
fermionic currents $G^{\pm}(z)$. Thus the coupling of the charge $\eta$ to the
$SO(2)$ Chern-Simons gauge field (representing the fermionic sector of the
$N=2$ superconformal field theory) effectively maps the representation of the
super-Virasoro algebra for $\eta=0$ to another representation with $\eta\neq0$
via a unitary transformation. This is the spectral flow, and it expresses the
fact that the $N=2$ superconformal algebras for different values of $\eta$ are
all isomorphic to each other. In particular, the spectral flow from the
Neveu-Schwarz sector ($\eta=0$) to the Ramond sector ($\eta=1/2$) has the
simple three-dimensional picture of adiabatically switching on the abelian
charge $\eta=1/2$ in the matter-coupled Chern-Simons gauge theory (\ref{Ieta}).

The fact that the $\eta$-twisted sector can be represented so simply here by
merely coupling another abelian charge to the topological field theory is
equivalent to the simple multiplication map (in the sense of the operator
product in conformal field theory) between representations of the $N=2$
superconformal algebra for different values of $\eta$. This latter mapping can
be represented as a simple shift of the $U(1)$ current $J(z)$ \cite{greene}.
Note that in general $J(z)$ can be represented in terms of an $SO(2)_2$ free
scalar boson field $X(z)$ (or equivalently a free complex fermion field
$\psi(z)$) as \cite{greene,aspinwall,ks,witten2}
\begin{equation}
\mbox{$J(z)=i\sqrt{\frac c3}~\partial_zX(z)=\sqrt{\frac
c3}~:\psi^\dagger(z)\psi(z):$}
\label{currentexpl}\end{equation}
Since $X(z)$ can be identified with the pure gauge boundary degree of freedom
of the $SO(2)_2$ Chern-Simons field $B$ on $\partial\cal M$, the representation
(\ref{currentexpl}) shows explicitly how the $U(1)$ charge $J(z)$ of the
two-dimensional conformal field theory corresponds to an abelian magnetic flux
(of $B$) in the Chern-Simons description. More precisely, since
$B_z(z)=\partial_zX(z)$ on a disc $D\subset\cal M$ (in a constant time slice),
using Stokes' theorem the flux integral (\ref{flux}) can be written as an
integral over a closed contour ${\cal C}(D)$ bounding the disc $D$ as
\begin{equation}
\Phi=\oint_{{\cal C}(D)}\frac{dz}{2\pi i}~B_z(z)=-i\sqrt{\frac3c}~\oint_{{\cal
C}(D)}\frac{dz}{2\pi i}~J(z)=\delta_{{\cal C}(D)}X
\end{equation}
which gives an explicit relation between the abelian flux $\Phi$ and the $U(1)$
charge enclosed by the conformal generator $J(z)$.

\newsection{The GSO Conditions}

The spectral flow with $\eta=1/2$ is the mathematical statement of the
existence of an isomorphism of $N=2$ super-Virasoro algebras which continuously
interpolates between the Neveu-Schwarz and Ramond sectors. From a physical
point of view, since the Neveu-Schwarz sector gives rise to spacetime bosons
and the Ramond sector yields spacetime fermions, spectral flow by half a
unit has the interpretation of a spacetime supersymmetry operator. In fact, it
is intimately related to the {\sc gso} projection \cite{gso} and also to
modular invariance \cite{seibwitt,genus} of superstring theory. These two
important features require the inclusion of both the Neveu-Schwarz and Ramond
sectors in the Hilbert space of the world-sheet theory. We now show how this
can be achieved dynamically using our three-dimensional approach. For this, we
consider the structure of the Hilbert space of the abelian topologically
massive gauge theory \cite{tmgt}
\begin{equation}
S_{\tiny TMGT}[B] = \int_{\cal M}\left(-\frac{1}{4e^2}F(B)\wedge\star F(B)+
\frac{\tilde k}{8\pi}B\wedge F(B)\right)
\label{tmgt}
\end{equation}
for the $SO(2)$ gauge field $B$, where for the time being we consider an
arbitrary Chern-Simons coefficient $\tilde k$ (later we will specify to the
desired $\tilde k=2$). Here $F(B)=dB$ is the field strength of $B$, and we take
the three-dimensional spacetime to be the product manifold ${\cal
M}=\Sigma^g\times{\bb R}^1$ with $\Sigma^g$ a compact Riemann surface of genus
$g$. The kinetic term for $B$ explicitly breaks the topological invariance of
the pure gauge theory. It is included for full generality because radiative
corrections by dynamical matter fields coupled to a Chern-Simons gauge field
induce a Maxwell term for it. Furthermore, its presence allows
for the construction of different string worldsheet actions, including the
action for the heterotic string, using the topological membrane approach to
string theory \cite{heterotic}, and it also enables one to vary the choice of
worldsheet complex structure in the induced conformal field theory on
$\Sigma^g$ via its coupling to the metric of $\cal M$ \cite{tm}.

The Gauss' law constraint in the Weyl gauge ($B_0=0$) is
\begin{equation}
\frac{1}{e^2}\partial^i\dot B_i +
\frac{\tilde k}{8\pi}\epsilon^{0ij}F(B)_{ij}=0~.
\label{constraint}
\end{equation}
At each fixed time $t\in{\bb R}^1$, we can write the 1-form $B=B_i(x)\,dx^i$ on
$\Sigma^g$ using the Hodge decomposition
\begin{equation}
B=d\xi+\delta\chi+a(t)
\label{hodge}\end{equation}
where $a$ is a harmonic 1-form on $\Sigma^g$, $da=\delta a=0$. Substituting
the decomposition (\ref{hodge}) into the action (\ref{tmgt}) and using the
constraint equation (\ref{constraint}), we find
\begin{equation}
S_{\tiny TMGT}[B] = S_{\tiny f}[\phi]+ S_{\tiny L}[a]
\end{equation}
where
\begin{equation}
S_{\tiny f}[\phi]= \int_{\cal M}d^3x~\mbox{$\frac{1}{2}$}
\left(\dot{\phi}^2-(\partial_i\phi)^2-M^2\phi^2\right)
\end{equation}
is the free particle action for the non-local scalar field $\phi\equiv
\sqrt{\partial^2/e^2}\,\chi$ of (topological) mass $M=\tilde ke^2/4\pi$.
The topological modes in (\ref{hodge}) propagate according to the quantum
mechanical Landau action \cite{landau}
\begin{equation}
S_{\tiny L}[a]=\int dt~\left(\frac{1}{2e^2}\dot{a}^2_i
-\frac{\tilde k}{8\pi}\epsilon^{0ij}a_i\dot{a}_j\right)
\label{landau}
\end{equation}
which describes the motion of a charged particle of mass $\mu = 1/e^2$  on the
plane ($a_1,a_2$) in a uniform magnetic field $\tilde B=\tilde k/4\pi$. The
mass
gap is $\tilde B/\mu= \tilde ke^2/4\pi=M$, which is precisely the mass of the
gauge boson. Note that ($a_1,a_2$) span the configuration space of the field
theory (\ref{tmgt}). However, if the quantum field theory is reduced to the
first Landau level, then the configuration space becomes the phase space of the
theory. A reduction to the first Landau level is reached by taking the limit
$\mu = 1/e^2 \to 0$. In that limit,  the full topologically massive gauge
theory (\ref{tmgt}) reduces to the pure Chern-Simons theory for $B$ which is an
exactly solvable three-dimensional topological field theory. In particular, the
first Landau level coincides with the moduli space of flat gauge connections on
the Riemann surface $\Sigma^g$ \cite{witten}.

The harmonic form $a$ can be written
\begin{equation}
a(t)=\sum_{p=1}^g\left(a_1^p(t)\alpha_p+a_2^p(t)\beta_p\right)
\label{harmdecomp}\end{equation}
in terms of a basis $\alpha_p,\beta_p$, $p=1,\dots,g$, of canonical harmonic
1-forms on $\Sigma^g$ which generate \mbox{$H^1(\Sigma^g;{\bb R})={\bb
R}^{2g}$}. These forms are normalized as the Poincar\'e duals
\begin{equation}
\oint_{a_{p'}}\alpha_p=\oint_{b_{p'}}\beta_p=\delta_p^{p'}~~~~~,~~~~~
\oint_{a_{p'}}\beta_p=\oint_{b_{p'}}\alpha_p=0
\end{equation}
to a basis $a_p,b_p$, $p=1,\dots,g$, of canonical homology cycles of
$\Sigma^g$, i.e. $a_p\cap b_{p'}=\delta_{pp'}$, $a_p\cap a_{p'}=b_p\cap
b_{p'}=0$, which generate its first homology group. After diagonalization of
the action (\ref{landau}) we get an independent copy of the Landau problem for
each pair $(a_1^p,a_2^p)$ of quantum mechanical topological modes from
(\ref{harmdecomp}), and so the full Hilbert space $\cal H$
of the abelian topologically massive gauge theory (\ref{tmgt}) is
\begin{equation}
{\cal H} = {\cal H}_{\tiny f}[\phi]\otimes\left({\cal H}_{\tiny
L}\right)^{\otimes g}
\end{equation}
where ${\cal H}_{\tiny f}[\phi]$ is the Hilbert space of the free massive
field $\phi$ and ${\cal H}_{\tiny L}$ is the Hilbert space
of the Landau problem on the plane.

If, however, the topologically massive
gauge theory is coupled to a (spectral) charge $q=-\eta$, then we actually get
$g$ copies of the Landau problem on the torus rather than the plane. To see
this, we consider the Wilson loop operators
\begin{equation}
W_{\cal C}^{(\eta)}[B] = \exp\left(i\eta\oint_{\cal C} B\right)
\label{wilson}
\end{equation}
which describe the holonomy that arises in adiabatic transport of the charge
with (closed) world-line $\cal C$ in the presence of the gauge field $B$. They
are invariant under the large gauge transformations
\mbox{($a_1^p,a_2^p)\rightarrow (a_1^p+2\pi m^p/\eta,a_2^p+2\pi n^p/\eta$)}
where $m^p,n^p$ are integers denoting the number of times the gauge field $B$
winds around the canonical homology cycles $(a_p,b_p)$ of $\Sigma^g$. Thus each
pair of quantum mechanical coordinates ($a_1^p,a_2^p$) lies on the torus $0\leq
(a_1^p,a_2^p)<2\pi/\eta$ of area $(2\pi)^2/\eta^2$. The density of states on
each Landau level is $\tilde B/2\pi$, and so the total number of states in each
level is ${\cal N}=(\tilde k/2\eta^2)^g$. Note that, for our purposes, the
topologically massive gauge theory (\ref{tmgt}) should be more precisely
defined  in terms of a matter-coupling to some minimal charge $\eta_{\tiny
min}$ with $\tilde k/2\eta_{\tiny min}^2\in{\bb Z}^+$. Then the Wilson loops
(\ref{wilson}), and hence the global gauge symmetries, must be defined in terms
of charges $\eta$ which are integer multiples of the fundamental one
$\eta_{\tiny min}$. Correlators of products of the Wilson loop operators
(\ref{wilson}) will then decompose into expectation values of products of the
basis ones $W_{\cal C}^{(\eta_{\tiny min})}[B]$. This construction partitions
the Hilbert space of the gauge theory (\ref{tmgt}) into superselection sectors
labelled by various choices of $\eta_{\tiny min}$, and the quantity $\tilde
k/2\eta_{\tiny min}^2$ in this way determines both the number of conformal
blocks and the chiral algebra of the respective conformal field theory on
$\Sigma^g$ \cite{landau}.

To describe the (topological) wavefunctions on the first Landau
level, it is convenient to use a holomorphic polarization in terms
of a complex structure of $\Sigma^g$. We write (\ref{harmdecomp}) as
\begin{equation}
a=\sum_{l=1}^{g}\left(\bar{a}^l\omega_l-a^l\bar{\omega}_l\right)
\end{equation}
where
\begin{equation}
\omega_l\equiv\alpha_l+\Omega_{lm}\beta_m
\end{equation}
are holomorphic harmonic 1-forms on $\Sigma^g$ and \mbox{$\Omega_{lm}$} is the
\mbox{$g\times g$} symmetric period matrix, with \mbox{Im\,$\Omega>0$},
which parametrizes the modular structure of $\Sigma^g$. The metric on the space
of holomorphic harmonic 1-forms is
\begin{equation}
G_{lm}\equiv i\int_{\Sigma^g}\omega_l\wedge\bar{\omega}_m=2\,
\mbox{Im}\,\Omega_{lm}
\end{equation}
Then the ${\cal N}=(\tilde k/2\eta^2)^g$ basis wavefunctions on the first
Landau level are \cite{wavefunc}
\begin{equation}
\Psi_{\{r_l\}}(a,\bar{a}|\Omega)=\exp\left(-\frac{\tilde k}{32\pi}a^l
G_{lm}\bar{a}^m\right)\exp\left(\frac{\tilde k}{32\pi}a^l G_{lm}
a^m\right)\Theta^{(g)}\left[
\begin{array}{c} 2r\eta^2/\tilde k \\ 0
\end{array}
\right]\left(\frac{\tilde ka}{4\pi\eta}\biggm|\frac{\tilde
k\Omega}{2\eta^2}\right)\label{Psi}
\end{equation}
where $r_l = 1,2,\ldots,\tilde k/2\eta^2$ and $l,m=1,\ldots,g$.
The Jacobi theta functions are defined by
\begin{equation}
\Theta^{(g)}\left[ \begin{array}{c} \alpha \\
\beta \end{array} \right] (z|\Pi) =
\sum_{\{n_l\}\in{\bb Z}^g} \exp\left[ i\pi(n_l+\alpha_l)\Pi_{lm}
(n_m+\alpha_m) + 2\pi i(n_l+\alpha_l)(z^l+\beta^l)
\right]\, ,
\label{theta}
\end{equation}
with the conditions $\alpha_l,\beta^l\in [0,1]$ and
Im\,$\Pi>0$ which ensure that $\Theta^{(g)}$ is a holomorphic function of
\mbox{$\{z_l\}\in {\bb C}^g$}. The wavefunctions $\Psi_r$ carry a
one-dimensional unitary representation of the discrete group of
large $U(1)$ gauge transformations
\begin{equation}
a^l~\rightarrow~a^l+s^l+\Omega^{lm}t_m~~~~~,~~~~~\bar{a}^l~\rightarrow~
\bar{a}^l+s^l+\bar{\Omega}^{lm}t_m
\end{equation}
where $s^l,t_m\in(2\pi/\eta){\bb Z}$,\, with
\begin{equation}
\Psi_r(a,\bar{a}|\Omega)~\rightarrow~\e^{-i\tilde
ks^mt_m/32\pi}\,\Psi_r(a,\bar{a}|\Omega)\, .
\end{equation}
The inner product of these wavefunctions is
\begin{equation}
\langle\Psi_r|\Psi_{r^\prime}\rangle\equiv\int_P\prod_{m=1}^g
da^m~d\bar{a}^m~\det
G^{-1}~\Psi_{r}^*(\bar{a},a|\bar\Omega)\Psi^{}_{r^\prime}(a,\bar a|\Omega)
=\delta_{rr^\prime}\det{\!}^{-1/2}G
\label{innerp}
\end{equation}
where, due to the large gauge invariance of this inner product,
the integration is restricted to the (phase space) plaquette
\begin{equation}
P=\left\{(a^l,\bar a^l)=u^l+(\Omega^{lm},\bar\Omega^{lm})v_m ~\Big|~ u^l,v_m
\in [0,2\pi/\eta)\right\}\, .
\end{equation}

We now consider the quantum mechanical mixing between a particular
linear combination of (vacuum) states in the first Landau level.
If we take the level $\tilde k=2$ and set $\eta=1/2$, then there are ${\cal
N}=4^g$ states in the first Landau level which are described by the
wavefunctions (\ref{Psi}) with $\{r_l\}=1,\ldots,4$. We will show that the {\sc
gso} projection corresponds to the following choice of basis wavefunctions on
the first Landau level,
\begin{equation}
{\cal L}=\left\{\frac{1}{\sqrt{2g\det{\!}^{-1/2}G}}
\sum_{l=1}^{g}\left(\Psi_{r_l}\pm\Psi_{s_l}\right) ~\biggm|~ (r_l,s_l)\in
\left\{(1,3),(2,4)\right\}\right\}\subset{\cal H}\, .
\label{gsobasis}
\end{equation}
Each state in $\cal L$ represents a particular spin structure on $\Sigma^g$
and quantum mechanical mixing between the ground states provides
a dynamical picture of modular invariance at the one-loop
level of superstring theory. To see this, we first recall that the gauge theory
(\ref{tmgt}) induces on $\Sigma^g$ the chiral {\em gauged} {\sc wznw} model (at
level $\tilde k$). The crucial observation is that the boundary boson fields
are minimally coupled to the $SO(2)_2$ gauge field and, upon
fermionization, we obtain gauged fermion fields (at level $\tilde k=2$)
on $\Sigma^g$. The fact that these fermion fields are gauged allows us to
encode their boundary conditions (i.e. spin structure) around each homology
cycle of $\Sigma^g$ in the following manner. When restricted to the first
Landau level, the fermionic terms in the (chiral) boundary action can be
written as
\begin{equation}
\bar\psi(z)(i\partial_z+B_z)\psi(z)=\bar\psi^\prime(z)~i\partial_z~
\psi^\prime(z)
\ \ \ \ \mbox{where}\ \ \ \
\psi^\prime(z)=\exp\left(i\int_{{\cal C}_z}B_z\right)\psi(z)
\label{bdryferm}\end{equation}
with ${\cal C}_z$ an oriented contour in $\Sigma^g$ from some fixed basepoint
to the point $z$. This means that the boundary fermion fields on $\Sigma^g$ can
be taken to be free, but with non-trivial parallel transport along the fibers
of
the spin bundle of $\Sigma^g$ determined by the holonomy of the $SO(2)$ gauge
connection $B$. From (\ref{bdryferm}) we see that the fermionic boundary
conditions
as one encircles homology cycles of the Riemann surface are in fact determined
by the vacuum expectation values of the Wilson loop operators (\ref{wilson})
around the canonical homology cycles $(a_p,b_p)$ of $\Sigma^g$, i.e.
\begin{eqnarray}
W_1^p&\equiv&W_{a_p}^{(1)}[B]~=~\exp\{ia_1^p\}~=~\exp\{i(\bar
a^p-a^p)\}\nonumber\\W_2^p&\equiv&
W_{b_p}^{(1)}[B]~=~\exp\{ia_2^p\}~=~\exp\{i(\Omega^{pl}\bar
a^l-\bar\Omega^{pl}a^l)\}
\label{wilsoncycles}\end{eqnarray}

The quantum mechanical amplitudes $\langle\Psi_r|W_i^p|\Psi_{r'}\rangle$ are
straightforward to calculate using the wavefunctions (\ref{Psi}) and the inner
product (\ref{innerp}). For the linear combinations of ground states given
in (\ref{gsobasis}) the averages are $\langle W_i^p\rangle=\pm 1$
corresponding to anti-periodic/periodic fermionic boundary conditions around
each canonical homology cycle of $\Sigma^g$. For example, in the case of the
torus ($g=1$) with $\Omega=i$, we obtain the four spin structures
\begin{center}
\begin{tabular}{|c|c|c|}
\hline
$\Psi$                & $\langle W_1\rangle$ & $\langle W_2\rangle$ \\
\hline
$|1\rangle+|3\rangle$ & $+$ & $-$  \\
$|1\rangle-|3\rangle$ & $-$ & $-$  \\
$|2\rangle+|4\rangle$ & $+$ & $+$  \\
$|2\rangle-|4\rangle$ & $-$ & $+$  \\
\hline
\end{tabular}
\end{center}
On higher-genus Riemann surfaces, we obtain $g$ copies of this structure for
each canonical pair $(a_p,b_p)$, so that the averages of the Wilson loops
(\ref{wilsoncycles}) in this way encode the $4^g$ spin structures of $\Sigma^g$
determined from $H^1(\Sigma^g;{\bb Z}_2)=({\bb Z}_2)^{2g}$. Since the $4^g$
ground states are degenerate, quantum mechanical mixing between these states
implies a mixing between the Neveu-Schwarz and Ramond boundary conditions for
the fermionic modes on the string worldsheet. The sum of amplitudes over all
spin structures of states, determined from the combinations of Chern-Simons
vacua in (\ref{gsobasis}), is then equivalent to projecting the trace in the
partition function onto those states, in each Hamiltonian sector of the
worldsheet theory on $\Sigma^g$, with eigenvalue $+1$ of the Klein operator
$(-1)^F$ (i.e. the states of even fermion number). This is the {\sc gso}
projection \cite{gso}. It ensures spacetime supersymmetry in superstring
theory, and it is in fact a general consequence of modular invariance of the
theory on a genus $g=1$ surface (i.e. at the one-loop level in superstring
theory) \cite{seibwitt}. This latter property follows immediately from the form
of the $g=1$ basis states represented in the table above, as then modular
transformations of the torus mix these spin structures by interchanging
homology cycles and hence map these particular states into each other. The
modular invariance of the vacuum sector of the gauge theory Hilbert space for
$g>1$ can
be attained by multiplying the states by appropriate phases and taking linear
superpositions of them as determined by the modular transformation properties
of the wavefunctions (\ref{Psi}) in the first Landau level \cite{wavefunc}.

Thus the coupling of the $SO(2)_2$ Chern-Simons gauge field $B$ to the spectral
charge $\eta=1/2$ gives the appropriate mixing of the Neveu-Schwarz and Ramond
sectors of the Hilbert space as required by both the {\sc gso} projection and
modular invariance. This is the crucial role played by the $SO(2)_2$ field in
the three-dimensional representation of the minimal $N=2$ superconformal field
theories, in that it yields the appropriate sector of the full Hilbert space of
the quantum field theory (\ref{abc}) in which to make the desired projections.
Note that (\ref{bdryferm}) is the three-dimensional analog of the
transformation in (\ref{N=1wznw}),(\ref{psi'g}), so that the possibility to
include these sectors is a manifestation of the three-dimensional
representation of the $N=2$ supersymmetry via the gauge field $B$. The above
Landau level picture suggests a dynamical origin for these sectors, and, in
particular, for $N=1$ spacetime supersymmetry (from the point of view of the
topological membrane approach), through the quantum mechanics of the Landau
problem.

Actually, as discussed in \cite{greene}, in order for the spectral
flow to be a symmetry for Neveu-Schwarz fields appearing in the partition
function for the $N=2$ superconformal quantum field theory, and furthermore for
the existence of the appropriate spacetime supersymmetry operator, the
conformal
field theory must be projected onto a spectrum with odd integral $U(1)$ charge
eigenvalues of $J(z)$ \cite{proj}. This projection (in the sense of orbifolding
to get conformal field theory quotients) can be done in two steps. First, one
can project onto integral $U(1)$ charges, which in the three-dimensional
description means adjusting the coupling of the abelian charges and
Chern-Simons gauge fields in such a way so as to ensure a magnetic flux
quantization condition $\Phi\in\bb Z$. Then one can apply a generalized {\sc
gso} projection onto odd integral charges, which in the topological field
theory picture above means selecting the appropriate basis of states in the
first Landau level of the abelian topologically massive gauge theory for $B$
(namely those states with eigenvalues $(-1)^\Phi=-1$). In this way we have a
well-defined dynamical procedure for building spacetime supersymmetric theories
from such three-dimensional topological field theories. It is intriguing that
odd-integer flux quantization, an ingredient of some condensed matter
applications of Chern-Simons gauge theory (most notably to the fractional
quantum Hall effect \cite{fqhe}), is a crucial part of this dynamical
construction. Such fluxes are precisely what is necessary for the
fermion-boson transmutation in a Chern-Simons theory realization of anyons
\cite{anyon} (particles with fractional exchange statistics), because they
produce an additional factor of $-1$ in the Aharonov-Bohm phases which maps
bosons into fermions and vice versa. The occurence of spacetime supersymmetry
in the conformal field theory thus appears as an anyonic symmetry of the
matter-coupled Chern-Simons gauge theory.

\newsection{Conclusions}

In this Paper we have presented a dynamical interpretation of the spectral flow
between the Ramond and Neveu-Schwarz sectors of the $N=2$ super-Virasoro
algebra. The dynamical picture is that of the vacuum sector of the quantum
mechanical Landau problem arising from the coupling of matter of charge
$\eta=1/2$ to an $SO(2)_2$ Chern-Simons gauge field (representing the fermionic
sector of the $N=2$ superconformal field theory). Then the quantum
mechanical mixing of an appropriate basis of states, representing the spin
structure for the spinor fields of the supersymmetric theory on a Riemann
surface, yields the necessary projection onto states required by both the {\sc
gso} projection and modular invariance. Within the topological membrane
approach to string theory, this suggests dynamical and geometrical origins for
$N=1$ spacetime supersymmetry.

The spectral flow with integer-valued $\eta^{-1}={\cal F}>2$ leads to an
interesting generalization of the above constructions. When such a charge is
coupled to the $SO(2)_2$ gauge field $B$ on ${\cal M}=\Sigma^g\times{\bb R}^1$,
there are ${\cal F}^{2g}$ degenerate states in the first Landau level of the
Hilbert space and $\cal F$ possible phases that the fermion fields can acquire
upon parallel transport around the canonical homology cycles of $\Sigma^g$.
Such spinor fields live on an $\cal F$-fold cover of the frame bundle of
$\Sigma^g$ (generalizing the spin bundle of the Riemann surface), and hence
combinations of the Chern-Simons vacua will admit quantum mechanical mixing
between sectors of the superconformal field theory generalizing the
Neveu-Schwarz and Ramond sectors (and determined from $H^1(\Sigma^g;{\bb
Z}_{\cal F})=({\bb Z}_{\cal F})^{2g}$). The appearence of such fractional
statistical phases in the boundary conditions for the superpartner fields is
one feature of the so-called fractional superstrings \cite{fracsusy}.

The features we have described in this Paper strictly speaking only apply to
the $N=2$ superconformal minimal models. Although it is unclear how to describe
general $N=2$ superconformal field theories using such three-dimensional
constructions (yet the general Kazama-Suzuki coset models (\ref{ks}) can be so
described), there are some simple generalizations of the above analyses. For
instance, we can consider a $U(1)^D$ topologically massive gauge theory
\begin{equation}
S_{\tiny TMGT}^{(D)}[{\cal A}]=-\frac{1}{4e^2}\int_{\cal
M}d^3x~\sqrt{h}~h^{\mu\nu}h^{\lambda\rho}F({\cal A})^I_{\mu\lambda}F({\cal
A})^I_{\nu\rho}+
\frac{k_{IJ}}{8\pi}\int_{\cal M}d^3x~\epsilon^{\mu\nu\lambda}
{\cal A}_\mu^I\partial_\nu{\cal A}_\lambda^J
\label{tmgtD}\end{equation}
where $I,J=1,\ldots,D$ and $h_{\mu\nu}$ is the metric of $\cal M$. The action
(\ref{tmgtD}) induces on the boundary of the 3-manifold $\cal M$ the linear
sigma-model \cite{linsig}
\begin{equation}
S_\sigma[X]= -\frac{1}{4\pi\alpha^\prime}\int_{\partial\cal M}d^2z~
\left(\sqrt{h}~h^{\alpha\beta}g_{IJ}\,\partial_\alpha X^I\partial_\beta X^J
+\epsilon^{\alpha\beta}B_{IJ}\,\partial_\alpha X^I\partial_\beta X^J\right)\, ,
\label{sigma}
\end{equation}
with the identification $k_{IJ}=4(g_{IJ}+B_{IJ})/\alpha^\prime$. Here $X^I$ is
the pure gauge part of the $U(1)$ gauge field ${\cal A}^I$ on the boundary
$\partial\cal M$, and $g_{IJ}$ and $B_{IJ}$ are the graviton and antisymmetric
tensor condensates, respectively. The antisymmetric $B$-term degree of freedom
is a topological instanton term in the field theory, and now the
three-dimensional gauge theory (\ref{tmgtD}) describes a simple conformal field
theory in a $D$-dimensional target space with constant metric.

{}From the three-dimensional point of view there are now two ways of adding
$N=2$ supersymmetry to the model (\ref{sigma}). One method is the direct
generalization of the $N=1$ construction of \cite{heterotic} to the $N=2$
supersymmetric abelian topologically massive gauge theory (by adding Dirac
rather than Majorana spinor terms to (\ref{tmgtD})). The resulting boundary
conformal field theory has $N=2$ supersymmetry (when both chiral and
anti-chiral sectors are included) and can be used for various string
constructions, as in \cite{heterotic}. The other approach is to add $D$
independent copies of another set of topologically massive gauge theories
(\ref{tmgt}) for $SO(2)_2$ fields ${\cal B}^I$ (i.e. with $\tilde
k_{IJ}=2\delta_{IJ}$). Then, via fermionization, the boundary boson
fields so induced effectively acquire Fermi statistics. In this second
approach, we can minimally couple each of the $D$ gauge fields ${\cal B}^I$ to
charged matter with $\eta_I=1/2$, and then the above picture of quantum
mechanical mixing between the $4^g$ states on the lowest-lying Landau level
also provides a three-dimensional description of the {\sc gso} projection and
modular
invariance in the linear sigma-model. Thus the dynamical interpretation of
spacetime supersymmetry can also be carried through for this simple
generalization to (constant metric) target space degrees of freedom. It would
be interesting to describe more complicated superconformal field theories with
$N=2$ supersymmetry, such as non-linear sigma-models in Calabi-Yau target
spaces or Landau-Ginsburg orbifold models, directly in terms of
three-dimensional topological field theory.

\bigskip

\noindent
{\bf Acknowledgements:} {\sc l.c.} gratefully acknowledges financial support
from the University of Canterbury, New Zealand. The work of {\sc r.j.s.} was
supported in part by the Natural Sciences and Engineering Research Council of
Canada.


\newpage

\begin{thebibliography}{99}

\bibitem{greene} B.R. Greene, Nucl. Phys. B (Proc. Suppl.) 41 (1995)
92.

\bibitem{aspinwall} P.S. Aspinwall, in Proc. Trieste Summer School in High
Energy Physics, eds. E. Gava, A. Masiero, K.S. Narain, S. Randjbar-Daemi and Q.
Shafi, World Scientific (1995).

\bibitem{specflow} A. Schwimmer and N. Seiberg, Phys. Lett. B 184
(1987) 191.

\bibitem{witten} E. Witten, Commun. Math. Phys. 121 (1989) 351.

\bibitem{mooreET} G. Moore and N. Seiberg, Phys. Lett. B 220 (1989) 422.

\bibitem{gko} P. Goddard, A. Kent and D. Olive, Commun. Math. Phys. 103
(1986) 105.

\bibitem{sakai} N. Sakai and Y. Tanii, Progr. Theor. Phys. 83 (1990) 968.

\bibitem{kogan1} I.I. Kogan, Phys. Lett. B 390 (1997) 189.

\bibitem{szaboET} G. Amelino-Camelia, I.I. Kogan and R.J. Szabo,
Nucl. Phys. B 480 (1996) 413.

\bibitem{gso} F. Gliozzi, J. Scherk and D. Olive, Nucl. Phys. B 122
(1977) 253.

\bibitem{tm} I.I. Kogan, Phys. Lett. B 231 (1989) 377;\\ S. Carlip and I.I.
Kogan, Phys. Rev. Lett. 64 (1990) 148; Mod. Phys. Lett. A 6 (1991) 171.

\bibitem{ks} Y. Kazama and H. Suzuki, Nucl. Phys. B 321 (1989) 232.

\bibitem{witten2} E. Witten, Commun. Math. Phys. 92 (1984) 455.

\bibitem{susyiso} P. Di Vecchia, V.G. Knizhnik, J.L. Petersen and P. Rossi,
Nucl. Phys. B 253 (1985) 701;\\ D. Nemeschansky and S. Yankielowicz, Phys. Rev.
Lett. 54 (1985) 620;\\ A.N. Redlich and H.J. Schnitzer, Phys. Lett. B 167
(1985) 315; \\ J. Fuchs, Nucl. Phys. B 286 (1987) 455.

\bibitem{polyakov} A.M. Polyakov, Mod. Phys. Lett. A 3 (1988) 325.

\bibitem{KZ} V.G. Knizhnik and A.B. Zamolodchikov, Nucl. Phys. B 247 (1984) 83.

\bibitem{anyon} F. Wilczek, Phys. Rev. Lett. 48 (1982) 1144, 1146; 49 (1982)
957;\\ F. Wilczek and A. Zee, Phys. Rev. Lett. 51 (1983) 2250;\\ D. Arovas, R.
Schrieffer, F. Wilczek and A. Zee, Nucl. Phys. B 251 [FS13] (1985) 117;\\ C.-H.
Tze and S. Nam, Ann. Phys. 193 (1989) 419.

\bibitem{seibwitt} N. Seiberg and E. Witten, Nucl. Phys. B 276 (1986) 272.

\bibitem{genus} L. Alvarez-Gaum\'e, G. Moore and C. Vafa, Commun. Math. Phys.
106 (1986) 1;\\ P. Ginsparg and C. Vafa, Nucl. Phys. B 289 (1987) 414.

\bibitem{tmgt} J.F. Schonfeld, Nucl. Phys. B 185 (1981) 157;\\
 S. Deser, R. Jackiw and S. Templeton, Ann. Phys. 140 (1982) 372.

\bibitem{heterotic} L. Cooper and I.I. Kogan, Phys. Lett. B 383 (1996)
271.

\bibitem{landau} I.I. Kogan and A.Yu. Morozov, Sov. Phys. JETP 61
(1985) 1;\\
I.I. Kogan, Comm. Nucl. Part. Phys. 19 (1990) 305; Intern. J. Mod. Phys. A 9
(1994) 3887;\\
G. Dunne, R. Jackiw and C.A. Trugenberger, Phys. Rev. D 41 (1990) 661;\\
 A.P. Polychronakos, Ann. Phys. 203 (1990) 231;\\
 X.G. Wen, Intern. J. Mod. Phys. B 4 (1990) 239.

\bibitem{wavefunc} M. Bos and V.P. Nair, Phys. Lett. B 223 (1989) 61;\\
M. Bergeron, D. Eliezer and G.W. Semenoff, Phys. Lett. B 311 (1993) 137.

\bibitem{proj} A. Sen, Nucl. Phys. B 278 (1987) 423;\\ T. Banks, L. Dixon, D.
Friedan and E. Martinec, Nucl. Phys. B 299 (1988) 613;\\ C. Vafa, Mod. Phys.
Lett. A 4 (1989) 1169.

\bibitem{fqhe} G.W. Semenoff and P. Sodano, Phys. Rev. Lett. 57 (1986) 1195;\\
R. Prange and S. Girvin, eds., The quantum Hall effect (Springer, Berlin,
1987);\\ T.H. Haansson, S. Kivelson and T.H. Zhang, Phys. Rev. Lett. 62 (1989)
82.

\bibitem{fracsusy} P.C. Argyres and S.-H.H. Tye, Phys. Rev. Lett. 67 (1991)
3339;\\ K.R. Dienes and S.-H.H. Tye, Nucl. Phys. B 376 (1992) 297;\\ P.C.
Argyres, E. Lyman and S.-H.H. Tye, Phys. Rev. D 46 (1992) 4533;\\ P.C. Argyres,
K.R. Dienes and S.-H.H. Tye, Commun. Math. Phys. 154 (1993) 471.

\bibitem{linsig} I.I. Kogan, Mod. Phys. Lett. A 6 (1991) 501;\\ L. Cooper, I.I.
Kogan and K.-M. Lee, hep-th/9611107, to appear in  Phys. Lett. B (1997).

\end{thebibliography}
\end{document}